\documentclass[12pt]{article}

\begin{document}

\title {
$${}$$
{\bf XX Heisenberg Spin Chain and}\\
{\bf an Example of Path Integral}\\
{\bf with ``Automorphic'' Boundary }\\
{\bf Conditions}
$${}$$}

\author{Cyril MALYSHEV\\
$${}$$
{\it V.A.Steklov Institute for Mathematics RAS,}\\
{\it St.-Petersburg Department,}\\
{\it Fontanka 27, St.Petersburg, RUSSIA}\\
E-mail: malyshev@pdmi.ras.ru}

\maketitle

\def \al{\alpha}
\def \be{\beta}
\def \ga{\gamma}
\def \dl{\delta}
\def \ep{\varepsilon}
\def \ze{\zeta}
\def \th{\theta}
\def \ka{\varkappa}
\def \la{\lambda}
\def \si{\sigma}
\def \ph{\varphi}
\def \om{\omega}
\def \Ga{\Gamma}
\def \Dl{\Delta}
\def \La{\Lambda}
\def \Si{\Sigma}
\def \Ph{\Phi}
\def \Om{\Omega}
\def \cA{\cal A}
\def \cB{\cal B}
\def \cC{\cal C}
\def \cD{\cal D}
\def \BC{I\!\!\!\! C}
\def \BD{I\!\!\!\! D}
\def \BZ{Z\!\!\!Z}
\def \IM{\Im}
\def \RE{\Re}
\def \1{^{-1}}
\def \cd{\partial}
\def \CD{{\cal D}}
\def \CA{{\cal A}}
\def \CM{{\cal M}}
\def \CL{{\cal L}}
\def \ld{\ldots}
\def \CU{{\cal U}}
\def \BQ{I\!\!Q\!\!I}
\def \bQ{\bold Q}
\def \vt{\vartheta}
\def \w{\widetilde}
\def \h{\widehat}
\def \d{\dagger}
\def \t{\times}
\def \l{\langle}
\def \r{\rangle}
\def \Tr{{\rm Tr}\,}
\def \tr{{\rm tr}\,}
\def \diag{{\rm diag}\,}
\def \Det{{\rm Det}\,}
\def \z{\zeta}

\rightline{\bf Dedicated to the memory of A.G.Izergin}
$$
{}$$
$$
{}$$
\begin{abstract}
\noindent
New representation for the generating function of correlators of 
third components of spins in the $XX$ Heisenberg spin chain is 
considered in the form given by the fermionic Gaussian path 
integrals. A part of the discrete anti-commuting integration 
variables is subjected to
``automorphic'' boundary conditions in respect of imaginary time.
The situation when only a part of the integration variables is
subjected to the unusual boundary conditions generalizes more
conventional ones when ``automorphic'' boundary conditions appear 
for all sites in the lattice spin models. The results of the
functional integration are expressed in the form of determinants
of the matrix operators. The generating function, as well as the
partition function of the model, are calculated by means of
zeta-regularization. Certain correlation functions at nonzero
temperature are obtained explicitly.
\end{abstract}
$$
{}
$$
\rightline{\bf hep-th/0204007}
\rightline{\bf PDMI preprint 07/2002    }

\newpage

\section{Introduction}

The correlation functions of the quantum models which can be solved
by the Bethe ansatz method [1] can be represented as the Fredholm
determinants of certain linear integral operators. One of such {\it
determinant} representations has been obtained in [2] for a simplest
equal-time correlator of one-dimensional model of ``impenetrable''
bosons, which are described by the quantum non-linear Schr\"odinger
equation with infinite coupling. The result [2] has been generalized
to the case of correlators with different time arguments [3], and
also to the case of the Heisenberg $XX$ spin chain [4]. The 
determinant representations of the correlation functions enable to 
deduce the integrable non-linear partial differential equations for 
the correlators [1, 5].

In its turn, path integration technique can be used to calculate
the correlation functions in various quantum models [6, 7]. A path
integration approach has been suggested in [8] to calculate the
generating function of correlators of third components of spins in
the $XX$ and $XY$ Heisenberg spin models on a cyclic chain. The path
integrals obtained in [8] for the $XX$-case lead to the answers in 
the determinant form [9]. Further, the determinant representations 
have been deduced in [10] for the equal-time temperature correlators 
of all components of spins in the anisotropic $XY$-chain. Multiple
integration over a set of the Grassmann coherent states (instead of 
the path integration) is used in [10]. Subsequent development of the
approach [10] can be found in [11]. It should be noticed that $XX$
model also still continues to attract attention [12], and
determinant representations for the correlators in integrable models
are also actively studied [13].

The idea to use path integration technique for obtaining certain
determinant formulas for $XX$ and $XY$ Heisenberg models looks
attractive. Therefore, another version of the path integration 
approach is developed in the given paper to represent both the 
partition function and the  generating function of correlators of 
third components of spins in the Heisenberg $XX$-model. The approach
proposed below is based on a technical note pointed out in Refs.[14] 
which are concerned with the index theory and supersymmetric
quantum mechanics. Path integration is used in [14] to evaluate 
traces of the corresponding operators, and for some cases the 
functional integration is defined in [14] over trajectories subjected 
to non-conventional ``automorphic'' boundary conditions with respect
of imaginary time.

Path integrals considered in the present paper are
also over a set of variables a part of which is subjected to
``automorphic'' boundary condition (while another part -- to 
usual requirements of a fermion/boson-type). It should be 
noticed that our path integral representations do not imply a
straightforward implementation of the proposal [14]: a special
restoration of invariance of the Lagrangian under shifts of
imaginary time by a period is required. The method of
zeta-regularization is used below to handle the determinants
obtained. The generating function, as well as the partition
function of the model, are calculated. Certain correlation
functions at nonzero temperature are obtained explicitly.
It is demonstrated that the path integration approach
proposed admits a considerable simplification and enough
transparency for the problem in question.

From a physical viewpoint, basic ideas of zeta-regularization
($\ze$-regularization) have been formulated in [15, 16]. In
mathematical literature, usage of
$\ze$-re\-g\-u\-l\-a\-r\-i\-z\-ation is usually traced to [17].
$\ze$-Regularization turned out to be rather useful in physics to
calculate, say, the instanton determinants [18, 19], the Casimir
energy on manifolds [20], as well as the axial and conformal 
anomalies [21]. One should be referred to [15--17, 22] for exposition 
of $\ze$-regularization.

The paper is organized as follows. Section 2 contains the
basic notations and introductory remarks. The representation for the
generating function of correlators of $\si_z^{(k)}$-operators
(and also for the partition function) in the form given by the
fermionic path integrals with ``automorphic'' boundary conditions is
obtained in Section 3 ($\si_z^{(k)}$ implies $\si_z$ at $k$th site).
The most important formulas of $\ze$-regularization are given in
Section 4. Moreover, the partition function of the model is 
calculated in Section 4 with the use of the generalized 
$\ze$-function in the series form. The generalized $\ze$-function in 
the form of a Mellin transform is defined in Section 5, and it is 
used to calculate the generating function of the correlators of 
spins. Differentiation of the integrals obtained with respect to a 
parameter is also considered in Section 5, and some specific 
correlators are calculated.
Discussion in Section 6 concludes the paper.

\section{Notations}

Let us consider the Heisenberg $XX$-model (which is an isotropic
limit of more general $XY$-model, [23]) on a chain of the length 
$M$ ($M$ is even). Let $Q(m)$ be the number of particles operator 
on the first $m$ sites of the chain $(m\le M)$. We shall calculate
$\exp(\al Q(m))$ averaged over the ground state of the model
(our notations, though conventional, correspond to [4, 10]),
$$
\l e^{\al Q(m)}\r\equiv \l\Phi_0\mid e^{\al
Q(m)}\mid\Phi_0\r,
$$
using the formula
$$ \l e^{\al Q(m)}\r=\frac{\Tr(e^{\al Q(m)}e^{-\be H_{XX}})}
{\Tr(e^{-\be H_{XX}})},\quad \al\in\BC,
\eqno(1)
$$
where $H_{XX}$ is the Hamiltonian, $\be$ is inverse temperature,
and $\Tr$ means trace of operator. The Hamiltonian of the model,
$H_{XX}$, will be taken in the fermionic representation [23, 24].
In fact, $H_{XX}$ considered by us appears after the
Jordan--Wigner transformation from initial Pauli spin
variables to the canonical fermionic variables subjected to the
anti-commutation relations:
$$
\{c_k,c_n\}=\{c^\d_k,c^\d_n\}=0,\quad
\{c_k,c^\d_n\}=\dl_{kn}.
$$
In these notations, $Q(m)\!\equiv\!  \sum^m_{k=1} c^\d_k c_k$.

It can be verified, that $H_{XX}$ commutes with the
total number of particles ${\cal N}\equiv Q(M)$, while the
parity operator $(-1)^{{\cal N}}$ anti-commutes with the canonical
variables,
$$
\{(-1)^{{\cal N}},c_n\}=\{(-1)^{{\cal N}}, c^\d_n\}=0,
$$
and it commutes with $H_{XX}$ since the latter is bilinear in
$c_n$, $c^\d_n$. Therefore, two projectors
$P^\pm=\frac12(1\pm(-1)^{{\cal N}})$ can be defined in such way,
that $H_{XX}$ can be written in the form [24]:
$$
\begin{array}{c}
H_{XX}=H^+P^++H^-P^-,\\[0.4cm]
\displaystyle{
H^\pm=-\frac12\sum^M_{n=1}(c^\d_n c_{n+1}+c^\d_{n+1}c_n)+h
{\cal N}-\frac{hM}2},
\end{array}
\eqno(2)
$$
where $h>0$ is an external magnetic field. The superscript
$^{\pm}$ in (2) implies the following boundary conditions at the
ends of the chain:
$$
\begin{array}{c}
c_{M+1}=-c_1,\quad
c^\d_{M+1}=-c^\d_1\quad {\rm for}\quad H^+,\\[0.4cm]
\displaystyle{
c_{M+1}=+c_1,\quad
c^\d_{M+1}=+c^\d_1\quad {\rm for}\quad H^-}.
\end{array}
\eqno(3)
$$

We shall calculate the required average (1), i.e.,
$$
G(m)\equiv\frac1Z\,\Tr\Bigl( e^{\al Q(m)}e^{-\be H}\Bigr)
$$
(the index $XX$ is omitted), using the formula [10]:
$$
G(m)=\frac1{2Z}
(G^+_FZ^+_F+G^-_FZ^-_F+G^+_BZ^+_B-G^-_BZ^-_B),
\eqno(4.1)
$$
where
$$
\begin{array}{rcl}
G^\pm_F Z^\pm_F & \equiv & \Tr\Bigl(
                e^{\al Q(m)}e^{-\be H^\pm}\Bigr),\\
[0.4cm]
\displaystyle{
G^\pm_B Z^\pm_B} & \displaystyle{\equiv} &
\displaystyle{ \Tr\Bigl(
         e^{\al Q(m)}(-1)^{\cal N}e^{-\be H^\pm}\Bigr),
}
\end{array}
\eqno(4.2)
$$
and $H^\pm$ are defined by (2), (3). By an analogy with (4),
the partition function $Z\equiv$ $\Tr(\exp(-\be H))$
can also be written in the form
$$
\begin{array}{c}
\displaystyle{
Z=\frac12(Z^+_F+Z^-_F+Z^+_B-Z^-_B),}
\\[0.4cm]
\displaystyle{
Z^\pm_F=\Tr(e^{-\be H^\pm}),\quad
Z^\pm_B=\Tr((-1)^{\cal N}e^{-\be H^\pm})
}
\end{array}
\eqno(5)
$$
((4) results in (5) at $\al\to0$).

The representations for $G^\pm_FZ^\pm_F$ and $G^\pm_BZ^\pm_B$
are obtained in [8] as the path integrals over the
Grassmann variables. As the result, the final answers in [8]
take the form of determinants of finite-dimensional matrices. The
anti-commuting functional variables in [8], $\xi_n(\tau)$, turn out
to satisfy conventional (anti-)periodicity rules with respect of
imaginary time $\tau:\xi_n(\tau)=\pm\xi_n(\tau+\be)$ (at $n$th 
site).

The following observation can be found in Refs.[14], which are
concerned with the index theory and supersymmetric quantum 
mechanics. Let $a$, $a^\d$ be some fermionic canonical operators. 
Let us consider a unitary operator $Q_\vt\equiv\exp(i\vt a^\d a)$ of
$\CU(1)$-transformation, which acts on $a,a^\d$ as follows:
$$
Q_\vt aQ^\d_\vt=e^{-i\vt}a,\quad
Q_\vt a^\d Q^\d_\vt=e^{i\vt}a^\d.
$$
When calculating $\Tr(Q_\vt\exp(-\be H))$, it turns out to be quite
natural to come to a path integral over a variable subjected to
``automorphic'' boundary condition: $\xi(\tau)=-e^{i\vt}\xi(
\tau+\be)$. The latter implies that the integration variable is
transformed accordingly to a nontrivial representation of $\CU(1)$
when $\tau$ is shifted by the period $\be$. Let us note that another
example of ``automorphic'' boundary condition can be found in [25]
where spin $1$ and $1/2$ chain models are studied by the method of
functional integration.

It is not difficult to note that $\exp(\al Q(m))$ behaves
analogously:
$$
e^{\al Q(m)} c_ne^{-\al Q(m)}=
\left[
\begin{array} {rl}
e^{-\al}c_n, & 1\le n\le m\\[0.2cm]
        c_n, & m<n\le M,
\end{array}
\right.
\eqno(6)
$$
and one can use the idea [14] when calculating $G^\pm_F Z^\pm_F$,
$G^\pm_B Z^\pm_B$ (4.2).
The ``automorphic'' condition arises for all sites in the models
considered in [25]. It can be guessed that the peculiarity due to
(6) will be concerned with $m\le M$, and the ``automorphic''
condition should be expected to appear only for a part of sites.

To conclude, let us define the coherent states using the
f\-e\-r\-m\-i\-onic operators $c_n$, $c^\d_n$ which possess the Fock
vacuum $\mid0\r$:
$$
c_n\mid0\r=\l0 \mid c^\d_n=0,\quad
\forall n,\quad \l 0\mid 0\r=1.
$$
Namely,
$$
\begin{array}{c}
\displaystyle{
\mid x(n)\r=\exp
\left\{
\sum^M_{k=1} c^\d_k x_k(n)\right\}\mid 0\r
\equiv
\exp (c^\d x(n))\mid 0\r,}\\[0.5cm]
\displaystyle{
\l x^*(n)\mid  =\l 0\mid
\exp
\left\{
\sum^M_{k=1} x^*_k(n)c_k\right\}
\equiv
\l 0\mid \exp (x^*(n)c),}
\end{array}
\eqno(7)
$$
where $n$ is the discrete index running from 1 to $N$, and the
shorthand notations are used:
$\sum^M_{k=1}c^\d_k x_k\equiv c^\d x$, $\prod^M_{k=1}dx_k\equiv
dx$, etc. In fact, $N$ independent coherent states are defined
which are labeled by independent complex-valued Grassmann
parameters $x^*_k(n)$, $x_k(n)$. The following relations hold
for the states (7):
$$
c_k\mid x(n)\r=x_k(n)\mid x(n)\r,\quad
\l x^*(n)\mid c^\d_k=\l x^*(n)\mid x^*_k(n),
$$
$$
\l x^*(n)\mid x(n)\r=\exp (x^*(n)x(n)).
$$

\section{ The path integral}

Let us turn to the problem of rewriting (4) and (5) in a path
integral form. For definitness, let us consider
$$
G^\pm_F Z^\pm_F=
\int dz\,dz^* e^{z^*z}
\l z^*| e^{\al Q(m)} e^{-\be H^\pm}|z\r,
\eqno(8)
$$
where the trace of the operator is understood as the integral
over the anti-commuting variables [26], and the coherent states
$\l z^*|,|z\r$ are defined as follows:
$$
\l z^*|=\l 0|\exp(z^* c),\quad
|z\r=\exp(c^\d z)| 0\r.
$$
In order to go over to the path integral, let us divide
the interval $[0,\be]$ into $N$ parts of the length $\be/ N$,
and let us represent $\exp(-\be H^\pm)$ as a product of $N$ 
identical exponentials. Inserting $N$ completeness relations 
between the exponentials, let us transform (8) into
$$
G^\pm_F Z^\pm_F=\int dz\,dz^*
\prod^N_{n=1} dx^*(n) dx(n)\exp
\left(z^*z-\sum^N_{n=1} x^*(n)x(n)\right)
$$
$$
\eqno(9)
$$
$$
\times
\l z^*| e^{\al Q(m)}|x(1)\r \l x^*(1)|
e^{-\frac\be N H^\pm}|x(2)\r\ld
\l x^*(N)| e^{-\frac\be NH^\pm}|z\r,
$$
where $\mid x(n)\r$ and $\l x^*(n)\mid$ are defined in (7).

Using the properties of the coherent states we evaluate the
following entries:
$$
\l z^*| e^{\al Q(m)}| x(1)\r=
\exp
\left(
e^\al \sum^m_{k=1} z^*_k x_k(1)+\sum^M_{k=m+1} z^*_k x_k(1)
\right),
$$
$$
\l x^*(n)| e^{-\frac\be NH^\pm}| x(n+1)\r
{\stackrel{\simeq}{_{_{N\gg1}}}}
$$
$$
\simeq \exp
\left( x^*(n)x(n+1)-\frac\be N\,H^\pm (x^*(n),x(n+1))\right),
$$
$$
H^\pm(x^*(n),x(n+1)) \equiv
-\frac12 \sum^M_{k=1}(x^*_k(n)x_{k+1}(n+1)+x^*_{k+1}(n)x_k(n+1))
$$
$$
+h\sum^M_{k=1}x^*_k(n)x_k(n+1)-\frac{hM}2.
$$
Inserting them to (9), we obtain:
$$
G^\pm_F Z^\pm_F=\int dz\,dz^*
\prod^N_{n=1} dx^*(n)dx(n)\,\exp
\bigg\{
\sum^m_{k=1} z^*_k(z_k+e^\al x_k(1))
$$
$$
+\sum^M_{k=m+1}z^*_k(z_k+x_k(1))+x^*(1)(x(2)-x(1))+\ld +
x^*(N)(z-x(N))
$$
$$
-\frac\be N(H^\pm(x^*(1),x(2))+\ld+
H^\pm (x^*(N),z))
\bigg\}.
\eqno(10)
$$
Let us denote $x(N+1)\equiv z$, $x_k^*(0)\equiv e^\al z_k^*$
(for $1\le k\le m$) or $z_k^*$ (for $m<k\le M$),
and impose the conditions
$$
\begin{array} {cll}
x_k(0)=&-e^{-\al}x_k(N+1), & 1\le k\le m,\\[0.2cm]
x_k(0)=&-x_k(N+1),  & m<k\le M,
\end{array}
$$
and perform $N\to\infty$. As the result, R.H.S. of
(10) acquires the integral form:
$$
\int \prod_{\tau\in[0,\be]} dx^*(\tau)dx(\tau)\exp
\left(\int^\be_0 \CL(\tau)d\tau\right),
\eqno(11.1)
$$
where $\CL(\tau)$ denotes the Lagrangian:
$$
{\CL}(\tau)= x^*(\tau)\frac{dx}{d\tau}-H^\pm (x^*(\tau), x(\tau)),
\eqno(11.2)
$$
and the functional variables $x(\tau)$, $x^*(\tau)$ are subjected
to the conditions:
$$
\begin{array} {cll}
x_k(\tau)=&-e^{-\al}x_k(\tau+\be), & 1\le k\le m,
\\[0.2cm]
x_k(\tau)=&-x_k(\tau+\be), & m<k\le M.
\end{array}
\eqno(11.3)
$$
Generally speaking, the fields $x^*_k(\tau)$ are independent
integration variables. It is convenient to subject $x^*_k(\tau)$
to a requirement analogous to (11.3) but with $e^\al$ instead of
$e^{-\al}$.

The derivation of the representation (11) follows [14] strictly,
and it does not take into account the peculiar character
of our problem: the conditions (11.3) characterize two
independent sets of sites. In their turn, the Hamiltonians
$H^\pm$ with the nearest neighbour coupling are diagonal just in
the momentum representation. Therefore, the following circumstance
becomes essential (which is new in comparison with [14, 25]).

It can be assumed that a certain representation of the group of
shifts of $\tau$ by the period $\be$, i.e., $\tau\to\tau+\be$,
is defined by the traditional (anti-)periodicity rules
$x_k(\tau)=\pm x_k(\tau+\be)$, $k\in \{1, \ld, M\}$, as well as by
the conditions (11.3). The action $\int^\be_0\CL(\tau)d\tau$ in the
exponent of (11.1) is a well-defined object provided
$\CL(\tau)$ is invariant under the shifts of $\tau$. Such invariance
takes place for a conventinal boundary condition provided 
$\CL(\tau)$ is even in powers of the fields.

Let us use (11.3) to calculate the variation $\dl\CL(\tau)$:
$$
\dl\CL(\tau)=\CL(\tau+\be)-\CL(\tau)=
$$
$$
=\frac12 \Bigl[
(e^\al-1)\Bigl( x^*_{m+1}(\tau)x_m(\tau)+x^*_M(\tau)
x_{M+1}(\tau)\Bigr)+\Bigr.
$$
$$
\Bigl.
+(e^{-\al}-1)\Bigl( x^*_m(\tau)x_{m+1}(\tau)+
          x^*_{M+1}(\tau)x_M(\tau)\Bigr)\Bigr].
$$
The case of the origin of $\dl\CL(\tau)$ is simple:
the cyclic quadratic form
$$
\sum^M_{k=1}(x^*_k x_{k+1}+x^*_{k+1} x_k)
$$
is invariant under
$$
x_k\to \pm e^\al x_k,\quad
x^*_k\to\pm e^{-\al} x^*_k,
\eqno(12)
$$
provided $k\in\{1,\ld,M\}$ (the homogeneous ``gauge'' 
transformation), and it is not invariant provided (12) is valid only 
for a part of sites (the nonhomogeneous transformation). The rule 
(11.3) implies a nonhomogeneous representation of shifts
$\tau\to\tau+\be$, and, thus, the invariance turns out to be broken
for $\CL(\tau)$ (11.2). However, this symmetry can 
straightforwardly be restored as follows: one should replace 
$H^\pm(x^*(\tau),x(\tau))$ in the limiting formula (11.1) by another 
form of the following type:
$$
\w H^\pm(\tau)=-\frac12
\sum^M_{k=1}{}^{^\prime}(x^*_{k+1}(\tau)x_k(\tau)+ x^* _k
(\tau) x_{k+1}(\tau))
$$
$$
+h\sum^M_{k=1} x^*_k(\tau) x_k(\tau)
-\frac12
\Bigl[x^*_{m+1}(\tau)x_m(\tau) e^{-\frac\al\be\tau}+x^*_m(\tau)
x_{m+1}(\tau) e^{\frac\al\be\tau}
$$
$$
+x^*_{M+1}(\tau)x_M(\tau)
e^{\frac\al\be\tau}+x^*_M(\tau)x_{M+1}(\tau) e^{-\frac\al\be\tau}
\Bigr]-\frac{hM}2,
$$
where $\sum'$ means that the indices $k=m,M$ are omitted. The
Lagrangian
$$
\w{\CL}(\tau)=x^*(\tau)\frac{dx}{d\tau}-\w H^\pm(\tau)
$$
is invariant under $\tau\to\tau+\be$, the integration measure in
(11.1) is also invariant, and, finally, we obtain:
$$
G^\pm_FZ^\pm_F=\int \prod_{\tau\in[0,\be]} dx^*(\tau) dx(\tau)
\exp
\left(\int^\be_0\w{\CL}(\tau)d\tau\right).
\eqno(13)
$$

The main statement of the present paper reads that the representation
(13) (together with the conditions (11.3)) is a well-defined relation
which is alternative to the functional representation obtained in 
[8]. The actual calculation below is to argue this assertion. 
Equations (11.3) remind the definition of an automorphic function 
(automorphic form, [27]):
$$
g^* f\equiv f(gu)=r(g)f(u)\,,
$$
where $f(u)$ is an appropriate function (form), $g$ is an elememt
of a group of transformations acting on the argument $u$ (thus
generating an action of $g^*$ on $f$), and $r(g)$ denotes a
representation of $g^*$. It is why we can formally consider
(11.3) as ``authomorphic'' boundary conditions
to distinguish them from more conventional rules of
fermionic/bosonic (at $\al$ $=$ $i\pi k, k\in\BZ$) type.

Let us pass in (13) to the momentum representation:
$$
\begin{array}{rll}
\displaystyle{
x_k(\tau)=} &
\displaystyle{
(\be M)^{-1/2}\sum_p e^{i(\om\tau-i\frac\al\be\tau+qk)} x_p,} &
 1\le k\le m, \\ [0.4cm]
\displaystyle{
x_k(\tau)=} &
\displaystyle{
(\be M)^{-1/2}\sum_p e^{i(\om\tau+qk)} x_p, } &
m< k\le M,
\end{array}
\eqno(14.1)
$$
where $p=(\om,q)$, and the summation goes over the Matsubara
frequencies $\om=\pi T(2n+1),$ $n\in\BZ$, and over the
quasi-momenta $q\in X^\pm$. Two sets $X^\pm$,
$$
\begin{array} {rl}
X^+=&\{q=-\pi+\pi(2l-1)/M \mid l=1,\dots ,M \},\\[0.2cm]
X^-=&\{q=-\pi+2\pi l/M \mid l=1,\dots ,M \},
\end{array}
\eqno(14.2)
$$
correspond to two boundary conditions (3). Let us substitute (14.1)
into (13), rescale $x_p\exp(i(m+1)q/2)\to x_p$, and obtain:
$$
\begin{array}{c}
\displaystyle{
G^\pm_FZ^\pm_F=\int\prod_{\om_F,q\in X^\pm} dx^*_p dx_p \exp
(S^\pm(\al))},\\[0.5cm]
\displaystyle{
S^\pm(\al)=\sum_p(i\om-\ep_q)x^*_px_p+\frac\al\be
\sum_{\om_F, q, q'}
         Q^{(0)}_{qq'} x^*_{\om q}x_{\om q'}+\frac{Mh\be}2},
\end{array}
\eqno(15.1)
$$
where $\ep_q=h-\cos q$ is the band energy of the quasi-particles,
$$
Q^{(0)}_{qq'}=\frac1M\,\frac{\sin\frac m2(q-q')}
{\sin\frac{q-q'}2},
\eqno(15.2)
$$
$\om_F$ denotes summation over the fermionic frequencies, and
$q,q'\in X^+$ or $X^-$.

Let use the known formula for multiple Grassmann integrals [26],
$$
\int dx^* dx\exp (-x^*\CM x) =\det \CM,
$$
to obtain the following formal answers:
$$
G^\pm_FZ^\pm_F=e^{Mh\be/2}\Det
\left[(-i\om_F+\ep_q)\dl_{pp'}-\frac\al\be\,
\dl_{\om\om'}Q^{(0)}_{qq'}\right],
\eqno(16.1)
$$
$$
G^\pm_BZ^\pm_B=e^{Mh\be/2}\Det
\left[(-i\om_B+\ep_q)\dl_{pp'}-\frac\al\be\,
\dl_{\om\om'}Q^{(0)}_{qq'}\right],
\eqno(16.2)
$$
and
$$
Z^\pm_F=e^{Mh\be/2}\Det[(i\om_F-\ep_q)\dl_{pp'}],
\eqno(17.1)
$$
$$
Z^\pm_B=e^{Mh\be/2}\Det[(i\om_B-\ep_q)\dl_{pp'}],
\eqno(17.2)
$$
where $\om_F$ and $\om_B$ are the fermionic and bosonic frequencies.
The symbol $\Det$ denotes determinants of ``infinite-dimensional''
matrices, while $`\det$' is reserved for conventional matrices. The
modification of all the calculations for (16.2) and (17.2) is 
straightforward. It is convenient to denote the matrix operators, 
which appear in (16) and (17), as $A(\al)\equiv A_\al$ and $A$, 
respectively.

\section{ Zeta-regularization.}

We shall use $\z$-regularization [15, 16, 22] in order to assign
meaning to the determinants (16), (17). Notice that, in princilpe,
the partition function of $XY$-model can be written with minor
(in comparison with (17)) modifications [10]:
$$
\begin{array}{l}
\displaystyle{
Z = \frac12 (Z^+_F+Z^-_F+Z^+_B-Z^-_B)
},\\[0.3cm]
\displaystyle{
Z^\pm_F\equiv \Tr(e^{-\be H^\pm_{XY}})=e^{-\be E^\pm_0}
\Det[(i\om_F-E_q)\dl_{pp'}]},\\[0.3cm]
\displaystyle{
Z^\pm_B\equiv \Tr((-1)^{\cal N}e^{-\be H^\pm_{XY}})=e^{-\be E^\pm_0}
\Det[(i\om_B-E_q)\dl_{pp'}]},
\end{array}
\eqno(18)
$$
where
$$
E^\pm_0\equiv -\frac12 \sum_{q\in X^\pm} E_q,\quad
E_q=(\ep^2_q+\ga^2\sin^2 q)^{1/2}.
$$
Thus, let us turn, for a generality, to regularization of (18)
instead of (17).

We shall begin with the introductory notes. Usually, a generalized
$\ze$-function is related to an elliptic operator. Precisely, let
$\CA$ be a non-negative elliptic operator of order $p>0$ on a
compact $d$-dimensional smooth manifold. Let its eigen-values 
$\la_n$ being ennumerated by the multi-index $n$. The series
$$
\ze(s\mid\CA)=\sum_{\la_n\ne0}(\la_n)^{-s},
\eqno(19)
$$
which is convergent at $\RE s>d/p$, defines the generalized
$\z$-function of the operator $\CA$. This series defines
$\ze(s\mid\CA)$ as the meromorphic function of the variable
$s\in\BC$, which can be analytically continued to $s=0$. The formal
relation
$$
\lim_{s\to0}\frac{d\ze}{ds}(s\mid\CA)=\lim_{s\to0}
\left[-\sum_{\la_n\ne0}\frac{\log\la_n}{(\la_n)^s}\right]=-\log
\left(\prod_{\la_n\ne0}\la_n\right)
$$
allows to define a regularized determinant of $\CA$ as follows:
$$
\log\Det \CA=-\ze'(0\mid\CA).
\eqno(20)
$$
The Riemann $\ze$-function,
$$
\ze(s)=\sum^\infty_{n=1}n^{-s},
\quad \RE s>1,
\eqno(21)
$$
and the generalized $\ze$-function,
$$
\ze(s,\al)=\sum^\infty_{n=0}(n+\al)^{-s},\quad
\al\ne0,-1,-2,\ld,
\eqno(22)
$$
which are meromorphic in $s$, have a simple pole at $s=1$ with 
residue 1, and possess a continuation at $s=0$ [28], can be formally
considered as particular cases of $\ze(s\mid\CA)$ (19). Notice that
$\ze\left(s,\frac12\right)$ is the Gurvitz $\ze$-function, and
$\ze(s,1)=\ze(s)$.

Starting with (19), one can represent $\ze(s\mid\CA)$ as a Mellin
transform:
$$
\ze(s\mid\CA)= \frac1{\Ga(s)}\int^\infty_{0}t^{s-1}
[\Tr(e^{-\CA t})-\dim(\ker\CA)]dt.
\eqno(23)
$$
The integral (23) is defined at sufficiently large positive
$\RE s$ (precisely, at $\RE s>d/p)$; for other $\RE s$
its analytic continuation is required. The formula (23)
can be related [21, 22] to the definition of $\Det\CA$
by means of the proper time regularization [29]:
$$
\log\frac{\Det\CA}{\Det\CA_0}=\Tr
\left[\int^\infty_0(e^{-\CA_0t}-e^{-\CA t})\frac{dt}t\right].
\eqno(24)
$$
The definitions by means of (20), (23), and by means of (24)
coincide up to an infinite additive constant.

Now one can pass to the calculation of (18). Let us define
the following series, which can be expresed through $\ze(s,\al)$
(22):
$$
\begin{array}{c}
\displaystyle{
\ze^\pm_F(s\mid A)\equiv \sum_{\om_F,q\in
X^\pm}(i\om_F-E_q)^{-s}}\\[0.5cm]
\displaystyle{
=\left(\frac{\be}{2\pi i}\right)^s
\sum_{q\in X^\pm}
\left[
\ze\left(
s,\frac12+i\frac{\be E_q}{2\pi}\right)+
(-1)^s\ze
\left(s,\frac12-i\frac{\be E_q}{2\pi}\right)\right]},
\end{array}
\eqno(25)
$$
$$
\begin{array}{c}
\displaystyle{
\ze^\pm_B(s\mid A)\equiv \sum_{\om_B,q\in
X^\pm}(i\om_B-E_q)^{-s}}\\[0.5cm]
\displaystyle{
=\left(\frac{\be}{2\pi i}\right)^s \sum_{q\in X^\pm}
\left[
\ze\left(
s,i\frac{\be E_q}{2\pi}\right)+
(-1)^s\ze
\left(s,-i\frac{\be E_q}{2\pi}\right)\right]-\sum_{q\in
X^\pm}(-E_q)^{-s}}.
\end{array}
\eqno(26)
$$
The analytic continuations for $\ze(s,z)$ are known [28]:
$$
\ze(0,z)=\frac12-z,\quad
\ze'(0,z)=\log\frac{\Ga(z)}{(2\pi)^{1/2}},
\eqno(27)
$$
and they lead to the following answers:
$$
\begin{array}{c}
\displaystyle{
-\lim_{s\to0}\frac d{ds}\,\ze^\pm_F(s\mid A)=\sum_{q\in
X^\pm}\log (1+e^{c\be E_q})},\\[0.5cm]
\displaystyle{
-\lim_{s\to0}\frac d{ds}\,\ze^\pm_B(s\mid A)=\sum_{q\in
X^\pm}\log (1-e^{c\be E_q})},
\end{array}
\eqno(28)
$$
where $c=\pm1$ due to an arbitrariness when differentiating
$(-1)^s$ $=$ $\exp(\pm i\pi s)$. The series $\ze^\pm_F(s\mid A)$
and $\ze^\pm_B(s\mid A)$ should be considered as the generalized
$\ze$-functions of the diagonal operators $A$ in the series form
(19).

Choosing $c=-1$, and combining (28) with (20), one obtains
the following relations of the $XY$-model [9, 10]:
$$
\begin{array}{c}
\displaystyle{
Z^\pm_F=e^{-\be E^\pm_0}\prod_{q\in X^\pm}(1+e^{-\be E_q})
=\prod_{q\in X^\pm}2\cosh\frac{\be E_q}2},\\[0.5cm]
\displaystyle{
Z^\pm_B=e^{-\be E^\pm_0}\prod_{q\in X^\pm}(1-e^{-\be E_q})
=\prod_{q\in X^\pm}2\sinh\frac{\be E_q}2}.
\end{array}
\eqno(29)
$$
The total partition function should be calculated accordingly
to (5), the free energy is $F=-(1/\be M)\log Z$, while arbitrariness
in the choice of $c$ does not influence the magnetization
$M_z=-\cd F/\cd h$ and the entropy $S=-\cd F/\cd T$. Specifically,
one gets in the thermodynamic limit:
$$
F=-\frac1{2\pi\be}\int^\pi_0\log (2(1+\cosh\be E_q))dq.
\eqno(30)
$$
All the formulas obtained can be reduced at $\ga\to 0$ to the
$XX$-model.

\section{ Determinats of the operators $A(\al)$}
\subsection{The regularization}

Thus, in the previous section we have defined $\ze$-functions of the
diagonal operators $A$ (17), (18) in the series form. Let us now use
(23) to calculate the regularized determinants of the non-diagonal
operators $A(\al)$ (16). For instance, let us calculate $G^\pm_F$
(16.1).

Let us begin with the formal integral
$$
\frac1{\Ga(s)}\int^\infty_0 t^{s-1}\Tr\left[
e^{(i\om_F-\h\ep+\frac\al\be\h Q)t}\right]dt,
\eqno(31)
$$
where $\h\ep$ and $\h Q$ imply the matrices in the momentum
space, $\diag\{\ep_q\}$ and $Q^{(0)}_{pq}$ (15.2), accordingly,
while the indices $p$, $q$ run independently over $X^+$ or $X^-$.
Convergence of the integral (31) at the upper bound is respected at
$h>h_c=1$ ($h_c$ is the critical magnetic field [4]).
Regularization is necessary at the lower bound.

Let us use the asymptotical relation
$$
\Tr\left[ e^{(i\om_F-\h \ep+\frac\al\be\h
Q)t}\right]\stackrel{\longrightarrow}{_{t \to\,0}}\phi_0,
$$
where $\phi_0\equiv\phi_0(A_\al)$ is an infinite constant equal
to $\Tr(\dl_{pp'})\equiv\sum_{\om_F}\tr(\h \dl)$ ($\h\dl$ is a unit
$M\times M$ matix). Let us define the function $\rho(t)$:
$$
\rho(t)\equiv\Tr\left[ e^{(i\om_F-\h \ep+\frac\al\be\h
Q)t}\right]-\phi_0,\quad
t\in[0,1[\,,
\eqno(32)
$$
and divide the integral (31) into two parts. We rewrite (31)
using (32) as follows:
$$
\frac1{\Ga(s)}\int^\infty_1
t^{s-1}\Tr
\left[ e^{(i\om_F-\h \ep+\frac\al\be\h
Q)t}\right] dt+
\frac1{\Ga(s)} \int^1_0 t^{s-1}\rho(t)dt+\frac{\phi_0}{s\Ga(s)}.
\eqno(33)
$$
The function $\rho(t)$ is a formal series in powers of $t^n$, 
$n\ge1$. Besides,
$$
\frac1{s\Ga(s)}\simeq 1+\ga s+o(s),\quad \ga=-\psi(1),
\eqno(34)
$$
where $\psi(z)=(d/dz)\log\Ga(z)$. Therefore, (33), which is regular
at $s\to0$, defines an analytic continuation of (31) at any $\RE
s\ge0$. It just can be considered as the definition of
$\ze^\pm_F(s\mid A_\al)$ in the right half-plane of $\BC\ni s$.

Let us now consider the constant $\phi_0$ and the coefficients
which define $\rho(t)$. These numbers turn out to be nonzero for
differential operators on manifolds [22]. In our situation all these
coefficients are infinite, but finite values can be assigned to them
by means of (21), (22).

First of all, using $\ze(0)=-\frac12$ one obtains:
$$
\phi_0=M\sum_{\BZ}1=M(2\ze(0)+1)=0,
$$
where ${\displaystyle \sum_{\BZ}}$ can equivalently be replaced by
${\displaystyle \sum_{\BZ+\frac12}}$, and $2\ze(0)+1$ --  by
$2\ze\left(0,\frac12\right)$ (``$\ze$-regularized measure'' of the
set $\BZ$ is zero). Further, the divergence of the coefficients at
the powers of $t$ in $\rho(t)$ is given by the divergent sums
$$
\sum_{n\in\BZ+\frac12}n^m.
$$
It is reasonable to consider such sums as zeros at $m=2k+1$, $k\in
\BZ^+$, since $\left(l+\frac12\right)^m$ are odd. If $m=2k$,
$k\in\BZ^+$, then
$$
\sum_{n\in\BZ+\frac12} n^m=2\sum^\infty_{l=0}
\left(l+\frac12\right)^m= 2\ze
\left(-2k,\frac12\right)=2\left(\frac1{2^{2k}}-1\right)\ze(-2k).
$$
But $\ze(-2k)=0$ at $k\ge1$. It can be concluded that
``$\ze$-regularized'' coefficients are zero for $A(\al)$, and, thus,
only the first term is relevant in (33).

Let use (34) to pass from (33) to the relation:
$$
-\lim_{s\to0}\frac d{ds}\ze^\pm_F (s\mid A_\al)
=
-\int^\infty_1\tr
\left[ e^{(-\h\ep+\frac\al\be\h Q)t}\right]
\left(\sum_{\om_F}e^{i\om_Ft}\right)\frac{dt}t-
\int^1_0 \rho(t)\frac{dt}t-\ga\phi_0,
\eqno(35)
$$
where the Poisson formula enables to sum up over
$\om_F$. Then, R.H.S. of (35) takes the form:
$$
-\tr\sum^\infty_{k=1}\frac{(-1)^k}k
\left( e^{-\be\h\ep+\al\h Q}\right)^k-\int^1_0
\rho(t)\frac{dt}t-\ga\phi_0
$$
$$
=
\log\det\left(1+
 e^{-\be\h\ep+\al\h Q}\right)
-\int^1_0\rho(t)\frac{dt}t-\ga\phi_0.
$$
Let us also take into account that $\h Q^2=\h Q$, and so
$e^{\al \h Q}-1=(e^\al-1)\h Q$. Therefore,
$$
G^\pm_F=\frac{\Det\left[
(i\om_F-\ep_q)\dl_{pp'}+\frac\al\be\,\dl_{\om\om'}
Q^{(0)}_{qq'}\right]}
{\Det[(i\om_F-\ep_q)\dl_{pp'}]}
=
\det\left[1+(e^\al-1)\h Q(1+e^{\be\h\ep})\1\right].
\eqno(36)
$$
Additional renormalization of $\rho(t)$ and $\phi_0$ (to zero, in
fact) is irrelevant for $G^\pm_F$, i.e., for the ratio of the
determinants. However, when $m=M$, the corresponding operator
$A_\al$ becomes diagonal since $\h Q$ becomes the unit matrix
$\h\dl$. In this case, we can consider $\ze$-function in the
series form (19). Transparent adjusting of (25)--(28) at $m=M$
gives the same answer as (35) with $\rho(t)$ and $\phi_0$ being
zero.

In an analogous way we obtain:
$$
G^\pm_B=\det\left[
1+(e^\al-1)\h Q(1-e^{\be\h\ep})\1\right].
$$
It should be pointed out that, say, $G^\pm_F$ (36) (i.e., the
ratio of two $\Det$'s) can be regularized in a way looking more
conventionally:
$$
G^\pm_F=\Det\left[\dl_{pp'}+\frac\al\be\,\frac{\dl_{\om\om'}
Q^{(0)}_{qq'}}{i\om_F-\ep_q}\right]\,.
\eqno(37)
$$
Using the series formula
$$
\log G^\pm_F (\al) = \Tr\sum_{k=1}^{\infty}
\frac{(-1)^{k-1}}k\,\left[\frac\al\be\,\frac{\dl_{\om\om'}
Q^{(0)}_{qq'}}{i\om_F-\ep_q}\right]^k
$$
to calculate (37) one can check that (36) and (37) have the same
numerical coefficients at the powers of $\al$.

\subsection{ Differentiation of the determinants}

It is necessary to differentiate the generating function (1) over
$\al$ at $\al=0$ when calculating the correlators of third 
components of spins (the operator of third component of spin, 
$\si^{(m)}_z$, is defined as $\si_z$ at $m$th site) as follows 
[1, 4]:
$$
\lim_{\al\to0}\frac{d^n}{d\al^n}\,G(m).
$$
In fact, with regard at (31), it is suffice to do only a first
differentiation. The other ones would occur as usual 
differentiations of matrices [30].

The operator in question, $A(\al)$, is linear in $\al$:
$A(\al)\equiv A_1+\al A_2$. Let us calculate the first derivative
of $\Det A(\al)$ using the formal integral (33):
$$
\frac{(d/d\al)\Det A(\al)}{\Det A(\al)}=-\frac d{d\al}
\left(\int^\infty_0\Tr(e^{A(\al)t})\frac{dt}t\right)
\eqno(38)
$$
(the regularization at $t\searrow 0$ is irrelevant for the
differentiation over the parameter). In the spirit of the
Ray--Singer--Schwarz lemma [31], we shall use in
(38) the following relation:
$$
\frac d{d\al}\Bigl(\Tr(e^{A(\al)t})\Bigr)
=t\frac d{dt}\,\Tr\Bigl(B(\al)e^{A(\al)t}\Bigr),\quad
B(\al)\equiv A_2 A\1(\al).
$$
Then, the integral over $t$ can be calculated, and one obtains:
$$
\frac{(d/d\al)\Det A_\al}{\Det A_\al}=\Tr B(\al)
$$
$$
=\Tr\left(
\frac{\h Q}\be(i\om_F-\h\ep+\frac\al\be\h Q)\1\right)=
\tr\Bigl(\h Q(1+e^{\be\h\ep-\al\h Q})\1\Bigr),
\eqno(39)
$$
where the Cauchy formula for matrices [30] is used to sum up over
the frequencies. Knowing (39), one can calculate all the
differentiations required:
$$
\lim_{\al\to0}\frac{(d/d\al)\Det A_\al}{\Det A_\al}=
\tr\Bigl(\h Q(1+e^{\be\h\ep})\1\Bigr),
$$
$$
\eqno(40)
$$
$$
\lim_{\al\to0}\frac{(d^2/d\al^2)\Det A_\al}{\Det A_\al}=
\tr\Bigl(\h Q(1+e^{\be\h\ep})\1\Bigr)+
$$
$$
+\tr^2\Bigl(\h Q(1+e^{\be\h\ep})\1\Bigr)-\tr
\Bigl(\h Q(1+e^{\be\h\ep})\1 \h Q(1+e^{\be\h\ep})\1\Bigr),
$$
etc.

To conclude the section, let us use the formulas obtained
to calculate the correlators $\l\si^{(m)}_z\r$ and
$\l\si^{(m+1)}_z\si^{(1)}_z\r$. To this end, let us rewrite (40):
$$
\l Q{(m)}\r=\frac mM\,\sum_q(1+e^{\be\ep_q})\1,
\eqno(41)
$$
$$
\l Q^2(m)\r=\frac mM\,\sum_q(1+e^{\be\ep_q})\1
+\frac{m^2}{M^2}
\left[
\sum_q(1+e^{\be\ep_q})\1\right]^2-
$$
$$
-\sum_{p,q}\Bigl(Q^{(0)}_{pq}\Bigr)^2(1+e^{\be\ep_p})\1
(1+\ep^{\be\ep_q})\1.
\eqno(42)
$$
One obtains from (41) in the thermodynamic limit:
$$
\si_z\equiv\l\si^{(m)}_z\r=1-\frac1\pi \int^\pi_{-\pi}
\frac {dq}{1+e^{\be\ep_q}},
\eqno(43)
$$
where the definitions
$$
\l\si^{(m)}_z\r=1-2\CD_1\l Q(m)\r
$$
and $\CD_1 f(m)=f(m)-f(m-1)$ are used. The result (43) agrees with
the magnetization $\CM_z=-\cd F/\cd h$, which is calculated by means
of $F$ (30). We obtain from (42) ($m>0$):
$$
\l\si^{(m+1)}_z\si^{(1)}_z\r=\si^2_z-\frac1{\pi^2}
\left|\int^\pi_{-\pi} \!\frac{e^{imq}}{1+e^{\be\ep_q}}
\,\,dq\right|^2,
$$
where the definitions
$$
\l\si^{(m+1)}_z\si^{(1)}_z\r=2\CD_2\l Q^2(m)\r
+2\si_z-1
$$
and $D_2f(m)=f(m+1)-2f(m)+f(m-1)$ are used. The answers obtained
reproduce correctly the result of [4], and, thus, they witness in
favour of the strategy chosen which stems from the functional
definition (11.3), (13).

\section{ Discussion}

The path integral representations for the partition function and
for the generating function of static correlators of third
components of spins in the $XX$ Heisenberg chain are obtained
in the present paper. The present paper is close to [14], where
a path integration with ``automorphic'' boundary conditions has been
used to calculate traces of some operators in the index theory, and 
to [25] where the partition functions of spin $1$ and $1/2$ chain
models have been also obtained in the form of path integrals over
variables subjected to ``automorphic'' boundary conditions. The
distinction between [14, 25] and the present paper
consists in the fact that the ``automorphic'' boundary conditions
appear only for a part of sites (several first ones) of the chain
model in question. It is interesting that the situation, when
only a part of the lattice variables is subjected to the unusual
boundary conditions in imaginary time, can successively be handled.

The path integrals in question are regularized by means of
zeta-regularization. It is demonstrated that, from a practical
viewpoint (i.e., if only differentiations over the parameter
$\al$ are needed), the formula for the first derivative of the
generating function can be obtained without regularization of the
Mellin integral at the lower bound.

The paper provides further development of the technical finds
discussed in [14] and [25], and it can be useful in practical
calculations for other models where vacuum averages (traces) of
operator exponentials similiar to our $\exp(\al Q(m))$ are of
importance.

\section*{Acknowledgement}

It is pleasure to thank N. M. Bogoliubov, F. Colomo, V. S. Kapitonov,
V. E. Korepin, P. P. Kulish, J.-M. Maillet, H.-J. Mikeska,
A. G. Pronko, N. A. Slavnov, and P. G. Zograf for interesting
discussions. The research has been supported in part by RFBR,
Grant No. 01-01-01045

\end{document}